# Probing the halo of Centaurus A: a merger dynamical model for the PN population

A. Mathieu [1,2], H. Dejonghe [1] and X. Hui [3]

[1] Universiteit Gent, Sterrenkundig Observatorium, Krijgslaan 281, B–9000 Gent, Belgium
[2] Observatoire de Strasbourg, 11 rue de l'université, F–67000 Strasbourg, France
[3] Astronomy Department, Boston University, Boston, MA 02215

**Abstract.**
Photometry and kinematics of the giant elliptical galaxy NGC 5128 (Centaurus A) based on planetary nebulae observations (Hui et al. 1995) are used to build dynamical models which allow us to infer the presence of a dark matter halo. To this end, we apply a Quadratic Programming method. Constant mass-to-light ratio models fail to reproduce the major axis velocity dispersion measurements at large radii: the profile of this kind of models falls off too steeply when compared to the observations, clearly suggesting the necessity of including a dark component in the halo. By assuming a mass-to-light ratio which is increasing with radius, the model satisfactorily matches the observations. The total mass for the best fit model is $\sim 4 \times 10^{11} M_\odot$ of which about 50% is dark matter. However, models with different total masses and dark halos are also consistent with the data; we estimate that the total mass of Cen A within 50 kpc may vary between $3 \times 10^{11} M_\odot$ and $5 \times 10^{11} M_\odot$. The best fit model consists of 75% of stars rotating around the short axis $z$ and 25% of stars rotating around the long axis $x$. Finally, the morphology of the projected velocity field is analyzed using Statler's classification criteria (Statler 1991). We find that the appearance of our velocity field is compatible with a type 'Nn' or 'Nd'.

## 1. Introduction

Although HI kinematics in normal spiral galaxies provide strong observational indication for the existence of unseen matter at large radii, unquestionable evidence of the presence of dark matter in elliptical galaxies is hard to obtain. Independent investigations, such as studies of X-ray emission (e.g., Sarazin 1987), kinematics of globular cluster systems (e.g., Harris 1991), HI observations (e.g., Schweizer, van Gorkom & Seitzer 1989), gravitational lensing statistics (e.g., Maoz & Rix 1993), and recently galaxy-galaxy gravitational lensing (e.g.,

*Send offprint requests to*: A. Mathieu (Sterrenkundig Observatorium)

Brainerd, Blandford & Smail 1995) and velocity profile measurements (Carollo et al. 1995), tend to favor the idea that some ellipticals do indeed possess massive dark halos. However, if it is well established that no or little amount of dark matter is present in the core region of ellipticals (see e.g., de Zeeuw & Franx 1991), decisive and clear-cut results are still limited by the scarcity of available data at large distances.

Radially extended and accurate kinematical measurements are required in order to study the halo dynamics and mass distribution of elliptical galaxies. Unfortunately, data are most easily obtained in the central bright areas of the galaxies, out to roughly one effective radius ($r_e$). Only with considerable effort can one reach beyond that radius (Saglia et al. 1993, Winsall 1991). On the contrary, planetary nebulae (PN) can be observed up to a large radius. Kinematical information of this stellar luminous tracer will undoubtedly further our understanding of galaxy dynamics and formation mechanisms. In particular, PN can be used as test particles to probe the halo of galaxies.

Among the various formation scenarios that have been proposed to account for the observed properties of ellipticals, the merger hypothesis gave a strong impetus to the study of galaxy evolution. Over the last twenty years, considerable effort has been devoted to investigating this idea, motivated by convincing examples of recently merged systems in individual galaxies such as NGC 5128 (for a review, see Barnes & Hernquist 1992). In this paper, we study the halo dynamics of NGC 5128 (Centaurus A) using the PN kinematical and photometric data taken from the work of Hui et al. 1995 (hereafter referred to as H95).

Numerical simulations of the merging of two identical galaxies consisting of a self-gravitating disk, bulge and halo show that ellipticals formed by major mergers rotate slowly in the core and exhibit rapid rotation in the outer regions (Hernquist 1993). Furthermore, these simulations form merger remnants displaying misalignments between their minor axes and their rotation axes. Faint features such as shells at large radii are also reproduced by the models, lending further credence to the merger scenario.

Unlike many ellipticals, Cen A is a galaxy which has a significant amount of rotation along both photometric major and minor axes. Since rotation about both the intrinsic short and long axes is allowed in a triaxial potential, the so-called minor axis rotation is generally believed to be a robust indicator of triaxiality. Thus Cen A appears to be a suitable candidate, as a merger remnant, to develop triaxial dynamical models that can be interpreted within the merger paradigm. In Sect. 2 we summarize the data we will use to build our models. The adopted geometry of Cen A is described in Sect. 3.

The determination of the mass distribution in ellipticals would certainly yield a major clue in the search for a theory of galaxy formation. As many ellipticals have similar luminosity profiles, self-consistency has usually been a basic assumption in building models of mass distribution. However, these constant mass-to-light ratio models, where stars provide all the light and mass, may be inadequate to mimic the observed features of real ellipticals if substantial amounts of mass in the outer region reside in an optically dark component like in spiral galaxies. After discussing our assumptions and the modeling technique in Sect. 4, we therefore proceed to build first constant mass-to-light ratio models in Sect. 5, and compare the predictions of the best fitting self-consistent model to the observations. We show that a dark matter halo is necessary to accommodate the observations. The logical consequence of this result is further investigated in Sect. 6, where we present models of Cen A with a dark halo component, including distribution functions. Our main conclusions are summarized in Sect. 7.

## 2. The data

### 2.1. The surface brightness distribution

The surface brightness distribution of Cen A is well represented by a de Vaucouleurs law between approximately $2'$ and $8'$ (Dufour et al. 1979). The galaxy's apparent ellipticity $\epsilon = 10(1 - \frac{b}{a})$ varies from $\epsilon = 0.7$ at $r = 2.6'$ to $\epsilon = 2.6$ at $r = 9'$. The surface brightness of the best fitting de Vaucouleurs profile to the major axis photometric data is tabulated vs. radius in tab. 1. The fit is determined by the data between $2'$ and $8'$. Following Hui et al. (1993), we assume a distance of 3.5 Mpc; thus 1 arcmin corresponds to 1.02 kpc. According to Dufour et al. (1979), $r_e = 5.18$ kpc. The values in tab. 1, which range from 0.07 to $\sim 22$ kpc, are computed from the fit.

### 2.2. The PN survey

The survey of H95 covers the entire galaxy out to a radius of 10 kpc but reaches as far as 20 kpc ($\sim 4r_e$) along the photometric major axis. The data set consists of approximately 400 PN positions, magnitudes and velocities.

The PN velocity field displays unambiguous signatures of a triaxial potential: the line of zero rotation and the line of maximum rotation of the galaxy are not perpendicular; moreover, as indicated in Fig. 1, the misalignment between the line of zero rotation and the photometric minor axis is about $39° \pm 10°$ (H95).

The major axis rotation curve and velocity dispersion profile have been constructed by folding the PN at the NE onto those at the SW side, and binning the PN in $2'$-intervals inside $10'$ and in $5'$-intervals at larger radii. For the minor axis, PN at the SE side were folded onto the NW, and binned in $2.5'$-intervals (H95).

PN major and minor axis rotation and velocity dispersion data are shown in tab. 2. Rotation along the major axis increases with radius up to 100 km/s at $\sim 7$ kpc and is flat at larger radii; the velocity dispersion declines gently from $\sim 140$ km/s at 3 kpc to $\sim 90$ km/s at 20 kpc. Minor axis rotation amounts to $\sim 50$ km/s between 5 and 10 kpc and the velocity dispersion profile is a shallow declining curve. The local $V/\sigma$ is approximately equal to 1 along the major axis and 0.5 along the minor axis.

**Table 1.** Surface brightness of the best fitting de Vaucouleurs profile to the major axis photometric data as a function of radius

| Radius (kpc) | B-magnitude |
|---|---|
| 0.07 | 17.04 |
| 0.24 | 18.07 |
| 0.40 | 18.57 |
| 0.59 | 19.03 |
| 0.78 | 19.36 |
| 0.94 | 19.62 |
| 1.11 | 19.85 |
| 1.29 | 20.06 |
| 1.46 | 20.25 |
| 1.63 | 20.42 |
| 2.75 | 21.29 |
| 4.79 | 22.35 |
| 7.14 | 23.20 |
| 9.08 | 23.76 |
| 12.75 | 24.61 |
| 17.85 | 25.52 |
| 22.13 | 26.15 |

**Fig. 1.** Projected configuration of Cen A on the sky

### 2.3. Other kinematical indicators

The rotation curve of the disk was obtained from $H\alpha$ emission (Nicholson, Bland-Hawthorn & Taylor 1992) and neutral hy-drogen 21-cm emission measurements (van Gorkom et al. 1990). The line of nodes of the disk coincides with the photometric minor axis (position angle 125°) in the outer region between ∼ 1.5 and 4.5 kpc and bends towards the photometric major axis in the very center. Nicholson et al. derived the $H\alpha$ rotation curve by fitting inclined rings to the gas velocity field observed in $H\alpha$ (Bland, Taylor & Atherton 1987). van Gorkom et al. fitted a rotation curve using the position-velocity map of the HI-emission along the major axis of the dust lane, assuming that Cen A is seen edge-on. Both studies obtain a peak velocity around 250 km/s. The circular velocity appears to be roughly constant out to 6 kpc.

**Table 2.** Major- and minor- axis rotation and velocity dispersion data

| No. of PN | Radius (kpc) | $V_{rot}$ (km/s) | $\sigma$ (km/s) |
|---|---|---|---|
| Major axis | | | |
| 31 | 1.6 ± 0.4 | -2 ± 22 | 123 ± 16 |
| 80 | 2.7 ± 0.6 | 48 ± 16 | 143 ± 11 |
| 31 | 4.7 ± 0.6 | 83 ± 25 | 131 ± 17 |
| 17 | 7.0 ± 0.6 | 102 ± 29 | 121 ± 21 |
| 11 | 8.9 ± 0.6 | 97 ± 35 | 115 ± 24 |
| 19 | 12.5 ± 1.6 | 85 ± 30 | 129 ± 21 |
| 12 | 17.5 ± 1.5 | 146 ± 26 | 91 ± 18 |
| 7 | 21.7 ± 1.6 | 100 ± 33 | 88 ± 24 |
| Minor axis | | | |
| 94 | 1.0 ± 0.6 | 22 ± 13 | 130 ± 10 |
| 32 | 3.8 ± 0.7 | 19 ± 20 | 111 ± 14 |
| 35 | 6.2 ± 0.8 | 53 ± 16 | 97 ± 12 |
| 16 | 8.4 ± 0.7 | 62 ± 29 | 117 ± 21 |
| 5 | 10.6 ± 0.5 | 30 ± 32 | 72 ± 23 |

Recent observations of neutral hydrogen associated with the optical shells of Cen A (Schiminovich et al. 1994) suggest the existence of a partial ring of HI at a radius of 15'. Its rotation axis is roughly perpendicular to the rotation axis of the HI disk. Assuming that the shells are circular and using the axis ratio of the shells, Schmininovich et al. derive an inclination of 60°-70° between the plane of the ring and the plane of the sky. However, as the gas only forms a partial ring, it has probably not yet relaxed and it is likely that it has not had enough time to settle into regular orbits. Using the position-velocity map, they obtain a circular velocity of 250 km/s. So, the rotation is likely be flat at least out to 15 kpc. They find that the mass within 15 kpc is about $2 \times 10^{11} M_\odot$ and the mass-to-light ratio $M/\mathcal{L} \sim 6$.

New measurements of CO(2-1) emission along the inner dust lane of Cen A in the inner region ($r < 70''$) have been used to study the kinematics and morphology of the CO molecular gas (Quillen et al. 1992). A warped disk model, consisting of a set of inclined rings, was fitted to the data and yielded a total mass of $\sim 1.1 \times 10^{11} M_\odot$ inside $\sim 5.3$ kpc. A warped disk model has also been fitted to infrared data of the dust lane in the bands J(1.2 $\mu$m), H(1.6 $\mu$m) and K(2.2 $\mu$m) (Quillen et al. 1993). The model reproduces the morphology of the IR data in the central region ($r < 70''$).

## 3. The geometry of Cen A

Little can be done in the exploration of the structure of elliptical galaxies without the determination of their intrinsic shape, which is a starting point to gain some insight into their internal dynamics. As elliptical galaxies are triaxial, the recovery of their intrinsic structure from photometry and kinematics is rather intricate, the deprojection being highly non-unique. The problem is compounded by the fact that often the orientation of the principal axes with respect to the line of sight is not known. However the presence of a dust lane is extremely useful to eliminate the degeneracy in the orientation. Since preferred planes exist for the settling of a gaseous disk in a triaxial galaxy (Merritt & de Zeeuw 1983), the aspect ratio of the disk provides additional constraints on the orientation of the galaxy. Thus H95 managed to estimate the axis ratios and the viewing angles of Cen A, using a combination of PN photometric and kinematical data together with information on the disk orientation.

**Fig. 2.** Intrinsic axes and orientation of Cen A

In order to unambiguously define the orientation angles we will use in this paper, we denote by $x$, $y$, $z$ the long, intermediate and short axes of an ellipsoid centered on Cen A. In that cartesian coordinate system, we define the usual spherical coordinates $(\theta, \varphi)$, where $\theta$ is the polar angle with the $z$-axis and $\varphi$ is the azimuthal angle in the $(x, y)$-plane (with respect to the $x$-axis). We choose the positive $x$-axis such that it points, in projection, at the SW side of the photometric major axis, while the positive $z$-axis points, in projection, towards the NW along the photometric minor axis. The estimated intrinsic axis ratios are $b/a = 0.9$ and $c/a = 0.8$ (H95), and the triaxiality parameter $T$, defined as

$$T = \frac{1 - (b/a)^2}{1 - (c/a)^2}, \qquad (1)$$

is equal to 0.4; so, the intrinsic shape of the galaxy is triaxial, rather spherical.

Let $x'$, $y'$, $z'$ be the cartesian axes of a coordinate system centered on Cen A such that the $z'$-direction is parallel to the line of sight and the photometric major axis coincides with the $x'$-axis.

The disk lies in the $(y, z)$-plane and has an inclination of 73° to the plane of the sky (Nicholson, Bland-Hawthorn & Taylor 1992). Hence the line of sight $z'$ is located at $\theta = 90°$ and $\varphi = 107°$ (H95). In other words, the line of sight $z'$ lies in the $(x, y)$-plane and offsets the intermediate axis $y$ by 17° as shown in Fig. 2. The angular momentum of the disk is pointing away from us, in the $-x$-direction.

## 4. The dynamical modeling of Cen A

### 4.1. The modeling assumptions

Since elliptical galaxies seem to share an apparent simplicity initially, a large body of theory has been developed that describes them as one-component stellar systems. However, this very idealized picture became inaccurate with detailed modern observations, and the discovery of a rich variety of previously unexpected features of obvious dynamical importance has complicated the simple picture. Dust-lane ellipticals like Cen A are attractive targets in this respect.

Compelling indications of mergers can be found in the dust-lane ellipticals in which two kinematically-distinct subsystems can be identified. Again Cen A is a typical prototype of such system, whose dynamics may be interpreted as resulting from two populations, one rotating about the short axis and one rotating about the long axis of the galaxy.

If we assume the merger hypothesis, rotation about more than one axis is almost inevitable. Modeling such systems requires a geometry that does not impose one preferred rotation axis. Therefore we simply assume a spherical potential for our modeling. However the spherical symmetry is adopted only for the potential. Our models will reproduce the flattened photometry, implying that what we will call a constant mass-to-light ratio model is only such along the major axis. Along the minor axis, the mass-to-light ratio will slightly increase as a function of radius. We will see later that the mass-to-light ratio in Cen A must actually be increasing with radius to account for the kinematical observations.

We will describe these subsystems with 2-integral models: we consider distribution functions that depend on the energy and on the projection of the angular momentum vector on a rotation axis. The rotation axes are the $x$-axis (i.e. the intrinsic long axis) and the $z$-axis (i.e. the intrinsic short axis). Very conveniently, the intrinsic long and short axes project onto the photometric major and minor axes.

### 4.2. The potential

We apply Bendinelli's deprojection method (Bendinelli 1991; Monnet, Bacon & Emsellem 1992) using multi-gaussian approximation to get the mass density distribution from its projection. Basically, one approximates the surface brightness profile with a sum of gaussians by means of a non-linear fitting algorithm. Then the deprojection, which reduces to a one-dimensional Abel inversion procedure, becomes straightforward and analytical.

We apply the method using the fit of a de Vaucouleurs law to the surface brightness data along the major axis, yielding the spherical spatial mass density. The spherical potential then follows readily from the numerical integration of the Poisson equation.

### 4.3. The QP technique

With the above assumptions, we search for triaxial dynamical models which could fit both the adopted photometry and the major and minor axes PN kinematics. We apply the QP method developed by Dejonghe (1989) to build our models. Briefly, we assume that the distribution function (DF) can be approximated by a linear combination of simple basis functions such as powers of the binding energy $E$ and of the projection $L_{\text{axis}}$ of the angular momentum vector $\mathbf{L}$ along an arbitrary axis.

We choose two different generic kinds of basis functions:

– distribution functions without rotation
$$f_{\alpha\beta} = E^\alpha L_{\text{axis}}^{2\beta} \qquad (2)$$

– distribution functions with rotation around an axis
$$\begin{cases} f_{\alpha\beta} = E^\alpha L_{\text{axis}}^{2\beta} & (\epsilon L_{\text{axis}} \geq 0) \\ f_{\alpha\beta} = 0 & \text{otherwise,} \end{cases} \qquad (3)$$

where $\epsilon = \pm 1$, depending on the sense of rotation about the appropriate axis.

In a triaxial potential, stars can rotate about both the intrinsic short axis $z$ and long axis $x$, and the resulting angular momentum vector lies in the plane defined by these two axes. So as to account for rotation about the $z$- and $x$-axes, we consider components including four different families of basis functions: $L_{\text{axis}}$ can be either $L_x$ or $L_z$, and each case comes in the two flavors described above. All these components constitute a library, which the QP program will use to pick components from. It does so by consecutively determining the component wich minimizes the errors, and adding that component to the final set. In that sense, the program determines the best linear combination of the library basis functions that fits the observations and produces a positive distribution function in phase space. The simple forms of the basis functions allow analytical expressions for the moments of the distribution function, which depend on the underlying gravitational potential.

The distribution function we construct must be positive definite in all of phase space. Because the DF depends on the binding energy $E$ and the angular momentum vector $\mathbf{L}$, the phase space is four-dimensional. One can reasonably restrict the region of phase space inside which the positivity of the DF will be tested by specifying a maximum radius far beyond the outermost data points. This translates into a lower limit for the binding energy and an upper limit for the modulus of the angular momentum, thus providing boundaries of the phase space region.

## 5. Constant mass-to-light ratio models

First we shall assume a constant mass-to-light ratio in constructing mass models. In this case, a strong constraint on the mass is provided by the $H_\alpha$ rotation curve through the familiar $GM(r) \sim rV_{\text{circ}}^2$, while the mass distribution is completely determined by the photometry.

We find a total mass of about $M_{\text{tot}} \sim 1.8 \times 10^{11} M_\odot$ within 50 kpc and a mass-to-light ratio $(M/\mathcal{L}_B)$ of $\sim 4.5$, assuming a blue luminosity $\mathcal{L}_B = 4.0 \times 10^{10} \mathcal{L}_\odot$. Fig. 3 shows the contours of the projected mass density of the model. The surface density of the QP model is a fairly good fit to the de Vaucouleurs photometry field in the area constrained by the data. Contours of the difference between the two are plotted in magnitude in Fig. 3.

**Fig. 3.** Left: major and minor-axis rotation curves and velocity dispersion profiles. The solid line represent the best fit model with a constant mass-to-light ratio. The dotted line is the circular velocity curve. The total mass is $M_{\rm tot} = 1.8 \times 10^{11} M_\odot$. Top right: contour map of the projected luminous mass density field of the best fit model with a constant mass-to-light ratio. The isocontours vary from 23 to 29 magnitude, with a step of 1 magnitude. Bottom right: contour map of the residuals between the photometry field of the QP best fit model and the de Vaucouleurs photometry field. The contours are -0.05, 0.05, 0.1, 0.15 and 0.2 magnitude. The negative contours are the dotted lines

Comparison of the model's rotation and velocity dispersion profiles to the data for both major and minor axes is shown in Fig. 3. The kinematical predictions of the model are roughly consistent with the observations in the inner region up to $\sim 10$ kpc, but no satisfactory fit could be obtained for the major axis velocity dispersion profile at large radii ($> 10$ kpc) where the model consistently underestimates the projected velocity dispersion. This mismatching clearly points to unseen matter in the outer region. The discrepancy may be solved by a model including a luminous and a dark component. Some indication of the improvement a dark halo may bring can already be found with the fit along the minor axis. As the potential is built spherical using the photometric major axis data and the mass density distribution is actually triaxial, the mass-to-light ratio is not strictly constant. If it is constant along the major axis, it increases a little along the photometric minor axis. Thus the models we nonetheless call self-consistent include a small amount of dark matter mainly along the minor axis. This explains why these models fit better along the minor axis than they do along the major axis. The mass distribution of the halo is rounder than that of the light.

Our negative result confirms the results of H95. There, a very simple self-consistent model was used to estimate the luminous mass. By applying the isotropic Jeans equation to the PN major axis rotation and velocity dispersion profiles, and adopting a Hernquist mass model (Hernquist 1990), H95 found that the total luminous mass was $1.6 \times 10^{11} M_\odot$ and the mass-to-light ratio was $M/\mathcal{L}_B \sim 4$. Moreover, they found that a constant mass-to-light ratio model could not fit the observed PN velocity dispersion profile. Similarly, our more sophisticated models, which provide a detailed picture of the dynamics of Cen A, fail to reproduce the PN velocity dispersion profile along the major axis, thereby strongly supporting the idea that a dark halo must indeed be present in Cen A.

## 6. Models including a dark halo component

Now we shall adopt an increasing mass-to-light ratio as a function of radius, so that the dark mass density is distributed continuously from the center of the galaxy to the outer halo.

**Fig. 4.** Left: major- and minor-axis rotation curves and velocity dispersion profiles. The solid line represents the best fit model including a dark halo component.The dotted line is the circular velocity curve. The total mass is $4 \times 10^{11} M_\odot$. Top right: contour map of the projected luminous mass density field of the best fit model with an increasing mass-to-light ratio as a function of radius. The isocontours vary from 23 to 29 magnitude, with a step of 1 magnitude. Bottom right: contour map of the residuals between the photometry field of the QP best fit model with a dark halo and the de Vaucouleurs photometry field. The contours are -0.14, -0.1, -0.05, 0.05, 0.1, 0.14 and 0.2 magnitude. The negative contours are the dotted lines

Then the luminous and dark mass densities contribute to the gravitational potential.

As very little is known about the possible shape of the halo dark mass density distribution in elliptical galaxies, we shall simply assume that the dark mass density distribution is the luminous mass density distribution multiplied by a function of radius.

In particular, we assume the following form for the total spherical mass density:

$$\rho(r) = [1 + D(r)]\rho_{lum}(r), \qquad (4)$$

where $\rho_{lum}(r)$ is the luminous mass density and $\rho_{lum}(r)D(r)$ is the dark mass density at radius $r$. We take for $D(r)$ an increasing function of the radius $r$, such as a power law with a cutoff radius $r_c$. Two parameters are needed to define a power law, the index $n$ and a multiplicative constant $\lambda$:

$$D(r) = \lambda \left(\frac{r}{r_c}\right)^n \quad 0 \leq r \leq r_c \qquad (5)$$

The cutoff radius, 50 kpc in our models, coincides approximately with a reasonable outer edge for the de Vaucouleurs profile. For $r > r_c$, we choose $\rho(r) \propto r^{-6}$. This falloff in the density is suffiently steep, so that the total mass beyond $r_c$ contributes only 20% of the total dynamical mass. There is no more mass outside 100 kpc. Obviously, the spherical gravitational potential is computed by integration of the Poisson equation using the total mass density.

### 6.1. Photometry, rotation curves and velocity dispersion profiles

Major-axis and minor-axis rotation curves and velocity dispersion profiles of the QP best fitting model including a dark component are shown in Fig. 4. In this model, we assume

$$D(r) = 12(\frac{r}{r_c})^2 \quad 0 < r \leq r_c, \qquad (6)$$

where the exponent 2 is arbitrarily chosen, but the constant 12 follows from the adopted total mass. This mass is more or less determined by the kinematical data, and amounts to $M_{\rm tot} \sim 4 \times 10^{11} M_\odot$ and $\sim 50\%$ of the total mass is in an optically dark form. The $M/\mathcal{L}$ ratio of the best fitting model vs. radius is plotted in Fig. 5.

**Fig. 5. a** Mass-to-light ratio as a function of radius. **b** Mass distributions as a function of radius: total mass (solid line), luminous mass (dotted line) and dark mass (short-dashed curve). Hence solid = dotted + short-dashed. The data are in the inner $\sim 20$ kpc

The dotted line in Fig. 4 is the circular velocity curve. The peak is about $\sim 250$ km/s at 4 kpc and the velocity decreases to $\sim 210$ km/s at 15 kpc. The major axis rotation curve of the model fits remarkably well the observations. The rotation peak is about 110 km/s at $\sim 6$ kpc and the rotation remains almost constant beyond 10 kpc at $\sim 100$ km/s. The minor axis rotation curve of the model is in good agreement with the observations, given the uncertainties in the data. The fit reaches a peak value of $\sim 60$ km/s at 3 kpc and then, the curve declines slowly at larger radii.

The fitted minor axis dispersion profile is consistent with the observations. Compared to the constant mass-to-light ratio model, the fit has quite improved in the inner region. The predicted velocity dispersion is constant or slightly decreasing beyond 15 kpc. Finally, the major axis dispersion of the best fitting model shows no significant departure from the data in the center up to the region of the most radially extended measurements.

The fit of the surface brightness for this model is better than the fit of the surface brightness for the constant mass-to-light ratio model. This can be seen in a qualitative way when comparing the photometry panels in Fig. 3 and Fig. 4. To make this statement somewhat more quantitative, we computed the cumulative distribution of the residuals for both models. We find that for the self-consistent model 90% of the pixels in the photometry field inside 35 kpc depart from the de Vaucouleurs photometry by less than 0.21 magnitude, whereas the latter number drops to only 0.14 magnitude for the model with a dark halo.

In the inner $\sim 10$ kpc, the self-consistent model and the model with a dark halo yield overall good fits to the photometry and kinematics. This is consistent with the fact that the mass within 10 kpc is about $1.1 \times 10^{11} M_\odot$ for both models. But the addition of a dark halo has substantially improved the fit of the major axis velocity dispersion for the two outermost data points. Therefore, the detection of a dark halo hinges in a crucial way on the presence and value of these two data points.

In order to estimate the range of mass within which consistency with the data is found, we constructed models with different functions $D(r)$ and total masses. With a function $D(r)$ that is a quadratic law as a function of radius, we estimate that models with a total mass ranging from $3.5 \times 10^{11}$ to $6.0 \times 10^{11} M_\odot$ give satisfactory fits to the photometric and kinematical data. This means that the difference between the surface brightness of the model and the de Vaucouleurs photometry remains smaller than $\sim 0.1$ magnitude almost everywhere in the region up to 35 kpc. In addition, the fit of the kinematics along the photometric major and minor axes is required to lie within the error bars, in particular for the velocity dispersion data at large radii. If masses are smaller than $\sim 3.5 \times 10^{11} M_\odot$, the fit generally underestimates photometry and kinematics, while the opposite is true for masses larger than $\sim 6.0 \times 10^{11} M_\odot$.

Models with slightly different viewing angles have also been investigated, so as to examine the consequence of uncertainties in the direction of the line of sight. In particular, if one assumes that the HI-ring observed by Schiminovich et al. (1994) has settled in a preferred plane of the potential, it implies that the line of sight may offset the $(x, y)$-plane by a small angle. By comparing the best fit models at different orientations, we found that a rather large range of viewing directions is compatible with the PN data: an offset of $\sim 20°$ in $\varphi$ or $\theta$ hardly influences the best fit of the PN photometric and kinematical data. No severe constraint can be put on the direction of the line of sight as we cannot discriminate among these best fit models.

The luminous mass of our model ($\sim 1.9 \times 10^{11} M_\odot$ inside 50 kpc) is consistent with that derived by van Gorkom et al. (1990) from HI-observation of the dust lane, which amounts to $1.75 \times 10^{11} M_\odot$. The simple spherical isotropic model of H95 leads to a mass distribution of Cen A that is not too different from our results. The dark and luminous mass density distributions are Hernquist mass models. The best fit model to the velocity dispersion profile has a total mass of $\sim 3.1 \times 10^{11} M_\odot$ within 25 kpc. The mass-to-blue-light ratio increases from 3.5 in the center to $\sim 10$ at 25 kpc. The mass-to-light ratios for our QP models increase from 4-5 at the center to 8-10 at 25 kpc and the total mass within 25 kpc is between $\sim 2 \times 10^{11} M_\odot$ and $\sim 3 \times 10^{11} M_\odot$ for models of total masses between $3.5 \times 10^{11} M_\odot$ and $6 \times 10^{11} M_\odot$.

Other estimates of the mass of Cen A have been given, based on different observations. Schiminovich et al. (1994) derived a mass of $2 \times 10^{11} M_\odot$ within 15 kpc from the observation of a HI-ring at that radius. For our models, the mass inside 15 kpc is about $1.5 \pm 0.1 \times 10^{11} M_\odot$, which is slightly lower. In contrast to these estimations, the total mass within $\sim 10$ kpc derived from X-ray observations is about $6 \times 10^{11} M_\odot$ (Forman, Jones & Tucker 1985).

### 6.2. The distribution function

The DF is a linear combination of basis functions which can be seen as a generalization of the functions used by Fricke

(1952). They depend on the energy and on the projection of the angular momentum vector along either the $x$-axis or the $z$-axis. We will call these functions $x$-rotators and $z$-rotators respectively. The total DF may be interpreted as resulting from the superposition of two subsystems, one rotating about the $z$-axis and the other about the $x$-axis.

Fig. 6 shows the sections of the DF in the planes $(E, L_z)$ and $(E, L_x)$. The values plotted are $-log(DF)$ except in the region where the DF is null: we arbitrarily set the corresponding value $-log(DF)$ to 0 in this region. The distribution function values are higher in the regions where $L_x$ is negative and $L_z$ is positive: this is consistent with the sense of effective rotation about the $x$- and $z$-axis. We choose basis functions of the form $f_{\alpha\beta} = E^\alpha L_{\text{axis}}^{2\beta}$ with $\beta > 0$. This implies that the DF is null when $L_{\text{axis}} = 0$.

Numerical simulations of merging encounters between equal-mass disk galaxies (Barnes 1992) provide a comparative starting point for an analysis of the orbital structure. Box orbits, X-tube and Z-tube orbits are supported by these simulated merger remnants. Minor-and major-axis rotation can be produced, with a resulting misalignment between the total angular momenta and the minor axes. Furthermore, Fig. 20b of Barnes (1992) shows the distribution of the Z-tube orbits of a remnant in the plane binding energy and $Z$-component of the angular momentum $(E, L_z)$. The global shape of the distribution resembles our section of the DF in the plane $(E, L_z)$. One can notice that the region where $L_z$ is null is clearly underpopulated, which is also the case for our model.

*6.3. The mass density distributions*

6.3.1. The projected mass density distributions

The moments of a Fricke-type component are analytical, so the moments of the DF are easily computed. Every star is an $x$-axis rotator or a $z$-axis rotator by design, so one can analyze separately the moments for the $x$-rotators and the $z$-rotators, which give rise to the X-tube and Z-tube orbits respectively. For example, one can determine what fraction of stars are rotating about the $x$- or $z$-axis by integrating the triaxial spatial mass density over all space for the $x$-rotators and for the $z$-rotators. We find that $\sim 25\%$ of stars are rotating about the $x$-axis (long axis) and $\sim 75\%$ about the $z$-axis (short axis) for the QP best fitting model with a dark halo.

One can also calculate the direction and modulus of the total angular momentum. The computation of the total angular momentum **L** reduces to the spatial integration of a simple linear expression involving moments of the distribution function. The angular momentum vector lies in the $(x, z)$-plane and its direction with respect to the $z$-axis decreases from $\sim 30°$ at 2 kpc to 14° at 50 kpc. It does not project onto the line of zero-rotation of the projected velocity field. The modulus of the total angular momentum for the $z$-rotators is approximately four times higher than the corresponding modulus for the $x$-rotators (Fig. 7).

The angular momentum can also be specified in terms of the usual dimensionless spin parameter $\lambda = |\mathbf{L}|E^{1/2}G^{-1}M^{-5/2}$. For our best fit model, $\lambda$ is approximately 0.06. It is small compared to the spin parameter of a self-gravitating disk which is $\sim 0.4$. Aarseth & Fall (1980) carried out N-body simulations of galaxy merging and found that the characteristic spin of merger remnants was about 0.07. More recently, N-body studies of mergers of equal-mass disc/halo

**Fig. 6.** 3D-plots of the distribution function: the axes are the binding energy which ranges from 0 to 1 and the angular momentum along the $z$-axis (top) and along the $x$-axis (bottom) (in $10^{13}$ $M_\odot.kpc.km/s$) and the vertical axis is $-log(DF)$. The DF is expressed in $\mathcal{L}_\odot/pc^3/(100km/s)^3$

**Fig. 7.** Top panels: total angular momentum (in $10^{13} M_\odot .kpc.km/s$) for the $x$-rotators (left) and for $z$-rotators (right) inside a sphere with a given radius. Bottom panels: the direction of the angular momentum with respect to the $z$-axis is plotted as a function of radius (left) and its direction (solid line) projected onto the plane of the sky is shown (right). The dashed line is the zero-velocity curve of the projected velocity field

galaxies (Barnes 1990) resulted in remnants exhibiting many properties of elliptical galaxies. Barnes considered encounters with different inclinations of the two incoming disks and found spin parameters in the range 0.04-0.13. Our value is consistent with both numerical experiments.

**Fig. 8.** Left: contour map of the projected luminous mass density for the $x$-rotators. The isocontours vary from 23 to 29 magnitude, with a step of 1 magnitude. Right: contour map of the projected luminous mass density for the $z$-rotators. The isocontours vary from 22 to 27 magnitude, with a step of 1 magnitude

Fig. 8 shows the contours of the projected mass densities for the subsystems rotating about the $x$- and $z$-axis. The projected mass densities are flattened in the direction of the angu-

**Fig. 9.** Contour map of the triaxial mass density (in magnitude) in the planes (top left) $x = 0$, (top right) $y = 0$ and (bottom left) $z = 0$. The step between the contours is 1 magnitude. Bottom right: the axis ratios are plotted as functions of the semi-major axis of the best fit ellipses to the isocontours of the spatial mass density

lar momentum of the components. The projected mass density for the total system is shown in Fig. 4.

### 6.3.2. The spatial mass density distributions

Fig. 9 shows the contours of the total spatial mass densities in the planes $x = 0$, $y = 0$ and $z = 0$. The values plotted are $-2.5 \times \log[\rho(\mathbf{r})]$ where $\rho(\mathbf{r})$ is the spatial triaxial mass density.

We fit ellipses to the contours and derive the axis ratios. They are plotted as a function of the semi-major axis of the ellipses. We find that this axis ratio of $x$-axis to the $y$-axis $a/b$ is about 0.95, and that the axis ratio of $z$-axis to the $x$-axis $c/a$ is approximately 0.7. The estimated intrinsic axis ratios in H95 are $b/a = 0.9$ and $c/a = 0.8$, which are not too different from the axis ratios of our resulting triaxial mass density. However the QP best fit model produces a triaxial mass density with the $y$-axis being the intrinsic long axis and the $z$-axis the intrinsic short axis. As the components we consider are purely axisymmetric $x$-rotators and $z$-rotators, it comes naturally that the mass density distributions are oblate. The $x$-rotators tend to produce a mass density flattened along the $x$-axis, and the intrinsic long axis turns out to be the $y$-axis. This is not a dynamical inconsistency however. Since our potential is spherical, every orbit is planar. What we call X-tubes and Z-tubes are in fact families of orbits with varying orbital planes such that the resulting density closely resembles the tubes found in a triaxial potential. In our case, a gas disk could settle in any plane, thus also in a plane that is forbidden in a triaxial potential. The QP model clearly shows that one should not necessarily identify the apparent long axis with the intrinsic long axis, as one is inclined to do. Our model nonetheless provides a realis-tic dynamical description of Cen A, but it makes it clear that a completely triaxial model of Cen A is also desirable.

*6.4. The velocity fields*

6.4.1. The projected velocity fields

The morphology of various velocity fields in elliptical galaxies has been studied by Statler (1991, hereafter referred to as S91). He produced a survey of projected velocity fields of triaxial, self-consistent, maximum entropy dynamical models, and then established a morphological classification based on features of the velocity fields at small and large radii.

Contours of the projected velocity field of the QP best fit model are plotted in Fig. 10. The velocity field of our model looks quite similar to the observed one (see H95). The line of zero rotation (the dashed line in Fig. 10) is inclined with respect to the photometric minor axis by approximately 30° in the central region. This in good agreement with the value derived by H95 from the observed velocity field, which is 39° ± 10°.

**Fig. 10.** Contour map of the projected velocity field of the best fit model with a dark halo. The isocontours vary from -110 to 110 km/s, with a step of 10 km/s. The negative contours are the dotted lines

The appearance of the projected velocity field of our model around and outside the effective radius displays a single maximum and minimum, and thus is consistent with an N-type or 'normal'-type, using the classification of S91.

Furthermore, three core types are defined in S91 depending on the deviation of the zero-velocity curve (ZVC) from the outer region to the center. If the deviation is less than 30°, the core is classified as $n$ or "normal"; if the deviation is between 30° and 90°, the core type is $d$ or "kinematically-distinct", and finally if the ZVC deviates by more than 90°, the core is "counter-rotating" or type $c$. The ZVC in our velocity field deviates by approximately 30° and thus the core could be either a $n$-type i.e. a "normal core" or a $d$-type i.e. a "kinematically-distinct core" in Statler's terminology.

**Fig. 11.** Left: contour map of the projected velocity field for the $x$-rotators of the best fit model with a dark halo. The isocontours are 0, ±30, ±50, ±70, ±90 and ±110 km/s. The negative contours are the dotted lines. Right: contour map of the projected velocity field for the $z$-rotators of the best fit model with a dark halo. The isocontours vary from -120 to 120 km/s, with a step of 20 km/s. The negative contours are the dotted lines

6.4.2. The spatial mean velocities

The moments of a distribution function constructed from Fricke-type components are analytical. Computation of the spatial velocity fields is therefore very easy. In a triaxial model, we need to consider three velocity fields, namely for the radial velocity $v_r$, the azimuthal velocity $v_\varphi$ and the polar velocity $v_\theta$ in spherical coordinates. Once again, we could calculate the velocity fields for the $x$-rotators on the one hand, and for the $z$-rotators on the other hand, so as to examine the contributions of the Z-tube and the X-tube orbits separately. The three-dimensional velocity fields of our QP triaxial model, as one could expect it, have a rather complex structure. So, we only present plane sections of the three spatial total velocity fields in three planes, perpendicular to the intrinsic axes $x$, $y$, $z$ and containing the center of the ellipsoid. Fig. 12 shows the sections of the azimuthal velocity $v_\varphi$ and of the polar velocity $v_\theta$ for the QP best fit model in the planes $x = 0$, $y = 0$ and $z = 0$. The radial velocity is zero.

In the plane $x = 0$, the azimuthal mean velocity is zero for the $x$-rotators, while it is positive for the $z$-rotators (with the direct sense of rotation around the $z$-axis). Hence we obtain a velocity field that is everywhere positive, and tends towards zero close to the rotation axis ($z$-axis). For the polar velocity, the situation is the opposite: the $x$-rotators (counter-rotating around the $x$-axis) show up with negative velocities in the semi-plane $y > 0$ and have a positive mean velocity in the semi-plane $y < 0$ (there is a discontinuity at $y = 0$ because of the definition of polar angle). The $z$-rotators have zero mean polar velocity.

In the plane $y = 0$, the azimuthal velocity of the $z$-rotators is positive. The mean azimuthal velocity of the $x$-rotators is negative in the quadrants $(x > 0, z < 0)$ and $(x < 0, z > 0)$, and is positive in the two other quadrants. The azimuthal velocity is discontinuous at $x = 0$. For $z$- and $x$-rotators, the polar velocity is zero in the plane $y = 0$.

In the plane $z = 0$, as expected for $x$-rotators (with the retrograde sense of rotation around the $x$-axis), the polar velocity is positive in the semi-plane $y < 0$ and negative in the semi-plane $y > 0$. The $z$-rotators do not contribute to the polar velocity field. The azimuthal velocity is zero for the $x$-rotators and is positive for the $z$-rotators.

**Fig. 12.** Top: sections of the azimuthal velocity $v_\varphi$ in the planes $x = 0$ (gray scale from $\sim 0$ to $\sim 130$ km/s) (left), $y = 0$ (gray scale from $\sim -125$ to $\sim 170$ km/s) (center) and $z = 0$ (gray scale from $\sim 40$ to $\sim 165$ km/s) (right). Bottom: sections of the polar velocity $v_\theta$ in the planes $x = 0$ (gray scale from $\sim -135$ to $\sim 135$ km/s) (left) and $z = 0$ (gray scale from $\sim -60$ to $\sim 60$ km/s) (right)

**Fig. 13.** Sections of $\sigma_r$ (top), $\sigma_\varphi$ (middle) and $\sigma_\theta$ (bottom) in the planes $x = 0$ (left), $y = 0$ (center) and $z = 0$ (right)

## 6.5. The velocity dispersions

### 6.5.1. The projected velocity dispersion field

Fig. 14 shows the contours of the projected velocity dispersion for the QP best fit model.

**Fig. 14.** Left: contour map of the projected velocity dispersion field for the QP best fit model. The contours vary from 120 in the center to 50 km/s in the outer region, with a step of 10 km/s. Right: azimuthal modulation of the velocity dispersion. The dotted line is the best fitting sinusoid curve to the data of H95. The solid line is the averaged velocity dispersion between 5 and 10 kpc for our QP model. The dashed-line is the best fit sinusoid curve to the solid line

Since the PN survey covers almost the entire galaxy to a radius of 10 kpc, one can examine the azimuthal variation of the velocity dispersion. To calculate the velocity dispersion, the PN between 5 and 10 kpc are binned in 15°-intervals (H95). We define $\Phi$ as the azimuthal angle on the sky with respect to the photometric minor axis and positive eastward. The velocity dispersion data are plotted as a function of the azimuthal angle in Fig. 14. H95 detected a correlation between the azimuthal modulations of the rotation and the velocity dispersion: the velocity dispersion and the rotation reach their maximum at the same azimuthal angle, and where the rotation is zero, the velocity dispersion is minimum. This is a clear indication that the observed rotation is not figure rotation. Furthermore, they fitted a sinusoid curve to the data:

$$\sigma = \sigma_m + \sigma_0 \sin(\Phi_0 + 2\Phi) \qquad (7)$$

The parameters of the best fit are: $\sigma_m = 111 \pm 6$ km/s, $\sigma_0 = 20 \pm 8$ km/s and $\Phi_0 = -169° \pm 23°$. In order to compare this fit with our best QP model, we compute the average of the projected velocity dispersion of the QP model between 5 and 10 kpc. The result is the solid line in Fig. 14. We obtain the following parameters for the best fit sinusoidal curve: $\sigma_m = 103$ km/s, $\sigma_0 = 19$ km/s, $\Phi_0 = -175°$. The fit is represented as the dashed line in Fig. 14. The amplitude, the period and the offset angle of the sinusoid are comparable to that of the best fit curve to the H95's data but there is a 10%-difference between the mean dispersions. Although our velocity dispersion is systematically lower than the fitted sinusoid to the data of H95, it remains almost everywhere within the error bars.

### 6.5.2. The spatial velocity dispersions

Fig. 13 shows the sections of the radial velocity dispersion $\sigma_r$ (top panels, gray scale from $\sim 75$ to $\sim 120$ km/s), sections of the azimuthal velocity dispersion $v_\varphi$ (middle panels, gray scale from $\sim 70$ to $\sim 170$ km/s) and sections of the polar velocity dispersion $v_\theta$ (bottom panels, gray scale from $\sim 70$ to $\sim 160$ km/s) for our best QP model in the planes $x = 0$ (left), $y = 0$ (center) and $z = 0$ (right).

The ratio of the cylindrical velocity dispersions $\sigma_z/\sigma_R$ varies from $\sim 0.5$ to $\sim 0.9$.

## 7. Conclusions

In this paper we construct triaxial dynamical models of NGC 5128 (Centaurus A) using the QP method on the basis of PN observations (H95). The PN kinematical and photometric data extend out to $\sim 4r_e$, and allow a detailed modeling of the halo dynamics of Cen A, an early type galaxy which probably has undergone a merger. The PN survey offers the opportunity to probe the halo of this nearby galaxy, and comparison with numerical simulations of merging systems shall likely shed some light on galaxy formation.

The QP modeling procedure is applied with the following assumptions. The models reproduce the photometry field of Cen A (classified E2), and rotation about both the short and the long axis is taken into account. The basis functions making up the distribution function are axisymmetric Fricke-type components, for which the symmetry axis can be the intrinsic short axis or the intrinsic long axis of the galaxy. The knowledge of the underlying gravitational potential is needed to calculate the moments of the distribution function. We simply adopt a spherical potential built *via* inversion of the major axis photometry, which is a rather realistic approximation. Finally, we consider self-consistent models and models with a dark matter halo component.

We find no self-consistent QP model with a constant mass-to-light ratio that can fit both major- and minor-axis rotation curves and velocity dispersion profiles together with the photometric data. While the QP model's photometry is in pretty good agreement with the data, the kinematical predictions of the models remain too far from the observations: in the central region, the velocity dispersion values of the QP models are substantially higher than the observations, and the models systematically underestimate the amount of velocity dispersion present on the major axis at large radii, as given by the two outermost data points. On the basis of the fit to these two points, the self-consistent model can be ruled out with 90% confidence.

On the contrary, a QP model including a dark halo can fit reasonably well the PN photometry and kinematics. The predictions of this model, compared to the self-consistent model, are globally as good or slightly better in the inner 10 kpc, but now the velocity dispersion profile beyond 10 kpc along the major axis provides a good fit to the data. The best fit model consists of $\sim 50\%$ of dark matter for a total mass of $4 \times 10^{11} M_\odot$. The mass-to-light ratio increases from 5 at 5 kpc to 12 at 50 kpc. Various dark matter models are possible, but in general we find that a reliable estimate of the total mass range interior to 50 kpc for Cen A is $3 \times 10^{11} M_\odot \leq M_{\rm tot} \leq 5 \times 10^{11} M_\odot$. This estimate based on stellar kinematics is compatible with the mass derived from HI-kinematics.

The axis ratios of the spatial mass density are $a/b \sim 0.95$ and $c/a \sim 0.7$. The QP best fit model consists of 75% of stars rotating about the short axis and 25% rotating about the long axis. The angle of the total angular momentum with respect to the short axis ($z$) decreases from $\sim 30°$ at 2 kpc to $14°$ at 50 kpc. The dimensionless spin parameter is about 0.06, which is of the same order of magnitude as the values derived from simulations of equal-mass disk/halo galaxy merging (Barnes 1990). Using Statler's morphological study of velocity fields (Statler 1991), we find that the velocity field of the best fit model can be classified 'Nn' or 'Nd'.

Our study shows that the kinematics and photometry of Cen A are well described by two kinematically-distinct subsystems rotating about the short and long axes of the galaxy respectively. Our model of Cen A shares many properties with the numerical simulations of merger remnants and it strengthens the idea that a merger scenario may play an important role in the formation process of dust-lane elliptical galaxies with minor axis rotation.

## 8. Acknowledgements

We would like to thank the anonymous referee for his comments, which helped us improve the paper.